\documentclass[pra,twocolumn,amsmath,longbibliography]{revtex4-1}

\usepackage{graphicx}
\usepackage{color}

\begin{document}

\title{Characterizing quasibound states and scattering resonances}

\author{Matthew D. Frye}

\affiliation{Joint Quantum Centre (JQC) Durham-Newcastle, Department of
Chemistry, Durham University, South Road, Durham DH1 3LE, United Kingdom}

\author{Jeremy M. Hutson}
\affiliation{Joint Quantum Centre (JQC) Durham-Newcastle, Department of
Chemistry, Durham University, South Road, Durham DH1 3LE, United Kingdom}

\date{\today}

\begin{abstract}
Characterizing quasibound states from coupled-channel scattering calculations
can be a laborious task, involving extensive manual iteration and fitting. We
present an automated procedure, based on the phase shift or S-matrix eigenphase sum, that reliably converges on a quasibound state
(or scattering resonance) from some distance away. It may be used for both
single-channel and multichannel scattering. It produces the energy and width of
the state and the phase shift of the background
scattering, and hence the lifetime of the state. It also allows extraction of
partial widths for decay to individual open channels. We demonstrate the method
on a very narrow state in the Van der Waals complex Ar--H$_2$, which decays
only by vibrational predissociation, and on near-threshold states of
$^{85}$Rb$_2$, whose lifetime varies over 4 orders of magnitude as a function
of magnetic field.
\end{abstract}

\maketitle

\section{Introduction}

Scattering resonances are important in many areas of physics
and chemistry. These include nuclear physics \cite{Feshbach:1962},
electron scattering from atoms \cite{Schultz:e-atom:1973} and molecules
\cite{Schultz:e-diatom:1973},
the spectroscopy of Van der Waals complexes \cite{Ashton:1983}
and chemical reaction dynamics \cite{Liu:2002}.
They have manifestations in both spectroscopy and collisions, and describe
both the decay properties of a quasibound state and the resonant scattering
that occurs at energies close to the state.

Zero-energy Feshbach resonances are particularly important in ultracold atomic
and molecular physics \cite{Chin:RMP:2010}. It is often possible to tune a
near-threshold state across a scattering threshold with an external (usually
magnetic) field. In this case the resonance properties are usually observed as
a function of field at essentially constant near-zero energy. Magnetically
tunable Feshbach resonances have been used both to control atomic interactions
via the scattering length and to create diatomic molecules by
magnetoassociation \cite{Kohler:RMP:2006}.

When there is a single open channel, the scattering phase shift $\delta(E)$
increases by $\pi$ across a resonance. Around an isolated narrow resonance it
follows the Breit-Wigner form as a function of energy,
\begin{equation}
\delta(E)=\delta_\textrm{bg}-\arctan\left(\frac{\Gamma/2}{E-E_\textrm{res}}\right),
\label{eq:BW}
\end{equation}
where $E_\textrm{res}$ is the resonance energy, $\Gamma$ is its width, and
$\delta_\textrm{bg}$ is the background phase shift. The lifetime of the
corresponding quasibound state is $\tau=\hbar/\Gamma$. In multichannel
scattering, the same behavior is shown by the eigenphase sum
\cite{Ashton:1983}, which is the sum of the phases of the eigenvalues of the
scattering S matrix, which all have magnitude 1. Individual S-matrix elements,
however, show more complicated behavior.

Scattering resonances are laborious to locate and characterize in scattering
calculations. It is usually necessary to carry out calculations on a grid of
energies across the resonance and fit the resulting phase shifts or eigenphase
sums to Eq.\ \eqref{eq:BW} \cite{Ashton:1983}. The width of the resonance is
not usually evident until quite late in the process, so many manual iterations
are often needed. In addition, the phase shift can usually be calculated only
modulo $\pi$, and it is easy to miss a narrow resonance entirely if the grid
used is too coarse. As an extreme example of this, some vibrational
predissociation resonances in the Van der Waals complexes Ar--H$_2$ and
Ar--D$_2$ are less than $10^{-9}$ cm$^{-1}$ wide \cite{Hutson:ArH2:1983}, and
occur at energies thousands of cm$^{-1}$ above some of their dissociation
channels. Locating such resonances in order to characterize them can be very
challenging.

The purpose of this paper is to describe an automated procedure for converging
on and characterizing resonances, using coupled-channel calculations of the
phase shift or S matrix. Once the location is approximately known, the procedure
can often characterize the resonance position, width and background phase shift
with calculations at fewer than 10 energies, without the need for manual
intervention. We will describe our method in terms of the phase shift, but it
applies equally to the eigenphase sum.

\section{Theory} \label{sec:theory}

The Breit-Wigner functional form, Eq.\ \eqref{eq:BW}, has three unknowns, and
thus we need the value of $\delta(E)$ at a minimum of three energies to
characterise the resonance. We first define the dimensionless quantity
$\widetilde{a}(E)=\tan \delta (E)$. This has a pole near the resonance position
and Eq.\ \eqref{eq:BW} can be rewritten as
\begin{align}
\widetilde{a}(E) &=\tan \delta_\textrm{bg} - \frac{(1+\tan^2\delta_\textrm{bg})\Gamma/2}
{E-E_\textrm{res}+(\Gamma/2)\tan \delta_\textrm{bg}}  \nonumber\\
&=\widetilde{a}_\textrm{bg}\left(1-\frac{\widetilde{\Delta}}
{E-\widetilde{E}_\textrm{res}}\right),
\label{eq:pole}
\end{align}
where
\begin{align}
\delta_\textrm{bg} &= \arctan \widetilde{a}_\textrm{bg};  \nonumber\\
\Gamma &=
2\widetilde{a}_\textrm{bg} \widetilde{\Delta} /
(1+\widetilde{a}_\textrm{bg}^2); \nonumber\\
E_\textrm{res} &=
\widetilde{E}_\textrm{res} + \widetilde{a}_\textrm{bg} \Gamma/2.
\label{eq:pole-to-BW}
\end{align}
The quantities $\widetilde{a}_\textrm{bg}$, $\widetilde{\Delta}$, and
$\widetilde{E}_\textrm{res}$ do not have immediate physical meaning, but they
put Eq.\ \eqref{eq:BW} into a form \eqref{eq:pole} that is more convenient
for numerical solution.

\begin{figure*}[tbp]
\centering
\includegraphics[height=0.85\textheight,clip=true,trim=2.0cm 1.9cm 1.5cm 1.8cm]{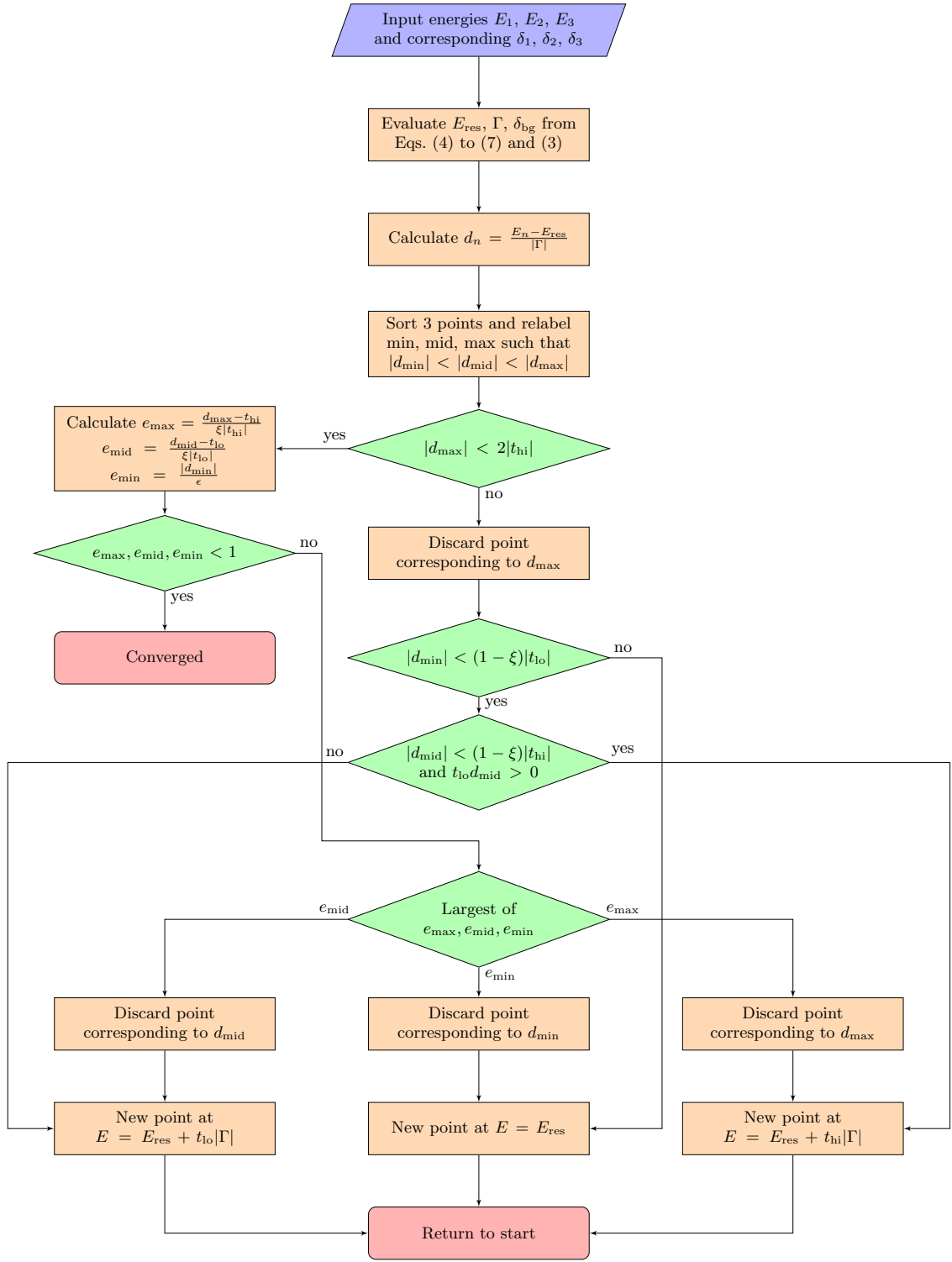}
\caption{Flowchart representation of the algorithm to select which point to
discard and where to place the next point. \label{fig:flowchart}}
\end{figure*}

Our numerical procedure is to evaluate the phase shift (or eigenphase sum)
$\delta(E)$ from coupled-channel scattering calculations at three energies
$E_1$, $E_2$ and $E_3$ to obtain the corresponding values $\widetilde{a}_1$,
$\widetilde{a}_2$ and $\widetilde{a}_3$. Solving 3 simultaneous equations
allows us to extract $\widetilde{a}_\textrm{bg}$, $\widetilde{\Delta}$, and
$\widetilde{E}_\textrm{res}$. Defining
\begin{equation}
\rho = \left(\frac{E_3-E_1}{E_2-E_1}\right)
\left(\frac{\widetilde{a}_2-\widetilde{a}_1}{\widetilde{a}_3-\widetilde{a}_1}\right),
\label{eq:rho}
\end{equation}
we obtain
\begin{align}
\widetilde{E}_\textrm{res} &= \frac{E_3-E_2\rho}{1-\rho} \label{eq:bres} \\
\widetilde{a}_{\textrm{bg}} \widetilde{\Delta} &=
\frac{(E_3-\widetilde{E}_{\textrm{res}})
(E_1-\widetilde{E}_{\textrm{res}})(\widetilde{a}_3-\widetilde{a}_1)}{E_3-E_1}
\end{align}
and finally
\begin{equation}
\widetilde{a}_{\textrm{bg}} = \widetilde{a}_1 + \frac{\widetilde{a}_{\textrm{bg}}
\widetilde{\Delta}}{E_1-\widetilde{E}_{\textrm{res}}}.
\label{eq:abg}
\end{equation}
Eqs.\ \eqref{eq:rho} to \eqref{eq:abg} and \eqref{eq:pole-to-BW} allow us to estimate the position,
width and background phase shift of the resonance from three points in its vicinity.
However, the estimates become numerically unstable if any two of the points are too close together. 
It is thus not satisfactory to converge on the resonance simply by generating a sequence of points
that approach closer and closer to $E_\textrm{res}$. Instead we aim to generate a final set of three 
points, one near $E_\textrm{res}$ and
two others at distances away from it comparable to the resonance width. In the present paper, we
converge the central point upon $E_\textrm{res}$. One of the outer points is placed about $\Gamma
t_\textrm{lo}$ from $E_\textrm{res}$ (with tolerance $\pm\xi\Gamma t_\textrm{lo}$), and the other
is about $\Gamma t_\textrm{hi}$ from it (with tolerance $\pm\xi\Gamma t_\textrm{hi}$). The three points can
be regarded as allowing characterization of $E_{\textrm{res}}$, $\Gamma$, and $\delta_\textrm{bg}$,
respectively. The logic we use to select which point to discard and where to place the next point
is shown in Fig.\ \ref{fig:flowchart}. We terminate the iteration when the estimated value of
$E_\textrm{res}$ is within a small amount $\epsilon$ of the closest of the 3 points and the other
two points satisfy the criteria above.

The algorithm shown in Fig.\ \ref{fig:flowchart} is similar to the one we presented previously for converging on a zero-energy Feshbach resonance in the scattering length as a function of external field \cite{Frye:resonance:2017}. However, it allows better control of the placement of the final points and improves the sequence of points close to convergence. It is beneficial in combination with the procedures of ref.\ \cite{Frye:resonance:2017} as well as for the present purpose.

The spacings $t_\textrm{lo}$ and $t_\textrm{hi}$ are signed, with $|t_\textrm{lo}|\le
|t_\textrm{hi}|$. The values $t_\textrm{lo}=-0.1$, $t_\textrm{hi}=1.0$ and $\xi=0.25$ are usually
appropriate for characterization of isolated resonances, and are used in the present work.
However, for special purposes, it may be necessary to use different choices of the points. Larger
values of $t_\textrm{lo}$ and $t_\textrm{hi}$ are sometimes needed for very narrow resonances, to
reduce the effects of numerical noise, and smaller values may be needed for very wide resonances,
to reduce variation in $\delta_\textrm{bg}$ across the range. For overlapping resonances, it may be
desirable to place both outer points on the same side of $E_\textrm{res}$.

We need three energies in the vicinity of the resonance to start the procedure.
We choose to use equally spaced points separated by a small amount $\delta E$,
based on a physical estimate of the resonance width. The initial estimate of
the resonance position could come from various sources, including approximate
calculations or experimental measurements; we usually use the program {\sc
bound} \cite{bound+field:2019, mbf-github:2020}, which solves the
coupled-channel problem subject to bound-state boundary conditions.

The algorithms described here make the approximation that
$\delta_{\textrm{bg}}(E)$ is constant across the range of points. This
approximation improves as the convergence proceeds and the range of points
becomes smaller. Nevertheless, it is the limiting factor that determines the
distance from which convergence can be achieved. At least one of the initial
points must give a phase shift that is affected by the resonance by more than
the variation of $\delta_{\textrm{bg}}(E)$ across the range of the points. For
very narrow resonances, computational noise in the phase shift can also affect
convergence.

\section{Numerical examples} \label{sec:total}

\subsection{Vibrational predissociation of Ar--H$_2$}

The first example we use to demonstrate this method is vibrational predissociation of the Van der
Waals complex Ar--H$_2$. The state with H$_2$ vibrational quantum number $v=1$, rotation $j=0$ and
total angular momentum $J=0$ lies about 4100 cm$^{-1}$ above the ground state and can predissociate
to form H$_2$, $v=0$, $j\le 8$. We use the interaction potential of Le Roy and Carley \cite{RJL80},
and perform close-coupled scattering calculations using the \textsc{molscat} package
\cite{molscat:2019, mbf-github:2020} to evaluate the S matrix and its eigenphase sum as a function
of energy. We use a space-fixed basis set that includes all functions with $j\le 10$ for $v=0$ and
with $j\le8$ for $v=1$ and solve the coupled equations using the symplectic log-derivative
propagator of Manolopoulos and Gray \cite{MG:symplectic:1995} with the six-step fifth-order
symplectic integrator of McLachlan and Atela \cite{MA5}.

This state has previously been characterized in coupled-channel calculations
\cite{Hutson:ArH2:1983}, but the purpose of this example is to show the
efficiency of the present method. We locate the resonance that describes this
state at $E=4139.075$ cm$^{-1}$ with respect to the threshold with $v=0$,
$j=0$. We find a width $\Gamma=2.02 \times 10^{-8}$ cm$^{-1}$, in agreement
with the result reported in Ref.\ \cite{Hutson:ArH2:1983}. The decay is very
slow because the interaction potential includes only low-order anisotropy and
the nearest open channel, with $j=8$, is only indirectly coupled to the bound
state. Open channels with lower $j$ are more directly coupled to the bound
state, but decay to them releases more kinetic energy and this reduces their
contributions.

\begin{table}[tbp]
\caption{Convergence on the $v=1$, $j=0$ state of Ar--H$_2$. Units of energy
are $10^{-10}\times hc$~cm$^{-1}$ and $E_\textrm{ref}=4139.075\times
hc$~cm$^{-1}$. The estimated value of $\delta_\textrm{bg}=0.486\pi$ remains
stable throughout the convergence. \label{table:ArH2}} \centering
\begin{ruledtabular}
\begin{tabular}{ccccc }
 & & & \multicolumn{2}{ c }{ Estimated values}  \\ \cline{4-5}
 $n$ & $E_n-E_\textrm{ref}$	& $\delta/\pi$		& $E_\textrm{res}-E_\textrm{ref}$	& $\Gamma$	\\ \hline
 $1$ & $2470000$ 			& $0.486303452$ 	& - 							& - 			\\
 $2$ & $2460000$ 			& $0.486303426$ 	& - 							& - 			\\
 $3$ & $2450000$ 			& $0.486303400$ 	& $1018699$ 					& $34.2$		\\
 $4$ & $1018699$ 			& $0.486291255$	& $369586$					& $72.0$		\\
 $5$ & $369586$			& $0.486258196$	& $-46283$					& $141.7$		\\
 $6$ & $-46283$			& $0.486009582$	& $-145463$					& $191.9$		\\
 $7$ & $-145463$			& $0.478484025$	& $-149553$					& $201.4$		\\
 $8$ & $-149553$			& $0.031357092$	& $-149567$ 					& $202.2$		\\
 $9$ & $-149587$			& $0.923863639$	& $-149567$					& $202.9$		\\
 $10$ & $-149364$			& $0.339217479$	& $-149567$					& $202.9$		\\
 $11$ & $-149567$			& $0.987000647$	& $-149567$					& $202.7$		\\
\end{tabular}
\end{ruledtabular}
\end{table}

Table \ref{table:ArH2} shows the sequence of points $E_n$ generated by our procedure
and how the successive estimates converge on the resonance parameters. Further details of the 
logic used at each step are given in the program output in Supplemental Material 
\footnote{See Supplemental Material at [URL will be inserted by publisher] for extracts of the \textsc{molscat} output file that describes the sequence shown in Table \ref{table:ArH2}.}. To demonstrate the power of the method, we have
purposefully picked an initial estimate that is further from the resonance than
the best estimates from bound-state calculations. Our method converges on this
extremely narrow resonance from over 10$^4$ widths away and characterizes it
using a total of only 11 calculations. In this case, as is usual, each point is based
on the three immediately preceding points. Most of the calculations
are used to converge on the resonance position $E_\textrm{res}$. Only
calculations 9 and 10 are placed away from the predicted resonance energy
before the final calculation is placed very close the resonance energy.

We have also characterized the same state using the improved
interaction potential of Le Roy and Hutson \cite{LeR87}. In this case we find
the resonance at $E=4138.884$ cm$^{-1}$, with width $\Gamma=9.4 \times
10^{-10}$ cm$^{-1}$, which is a factor of about 20 narrower than for the
potential of Le Roy and Carley \cite{RJL80}. The potential coupling terms
responsible for vibrational predissociation are anisotropic terms off-diagonal
in the H$_2$ vibrational quantum number, and these are dominated by the
coefficient $V_{\lambda k}(R)$ with $\lambda=2$ and $k=1$ in the potential
expansion of Refs.\ \cite{RJL80} and \cite{LeR87}. Figure 3 of Ref.\
\cite{LeR87} shows that this coefficient is significantly weaker for the
potential of Ref.\ \cite{LeR87} than for that of Ref.\ \cite{RJL80}, and this
is responsible for the decreased width.

\subsection{Lifetimes of $^{85}$Rb$_2$ Feshbach molecules}

The second example is for $^{85}$Rb$_2$ molecules produced by magnetoassociation at the $(f,m_f)$ =
$(2,-2) + (2,-2)$ threshold \cite{Thompson:spont:2005}. This is not the lowest threshold for
$^{85}$Rb atoms in a magnetic field, so the molecules can decay to lower thresholds in which one or
both of the atoms has $m_f>-2$. The Feshbach resonance used for magnetoassociation at a magnetic
field near 155~G is caused by a state that has a lifetime around 82 $\mu$s at energies well below
the $(2,-2) + (2,-2)$ threshold. As it approaches threshold, the bound state acquires a large
fraction of threshold character. The amplitude of its wavefunction at short range decreases in both
the closed channel and the threshold channel. This reduces its width $\Gamma$ and increases its
lifetime $\tau=\hbar/\Gamma$ close to threshold.

We use the interaction potential of Strauss \emph{et al.}\ \cite{Strauss:2010},
without retardation corrections. We perform coupled-channel calculations using
the \textsc{molscat} \cite{molscat:2019, mbf-github:2020} and \textsc{bound}
\cite{bound+field:2019, mbf-github:2020} packages. The methods used are as
described in Ref.\ \cite{Blackley:85Rb:2013}. The Hamiltonian includes atomic
hyperfine coupling, electron and nuclear Zeeman terms, and coupling due to the
singlet and triplet interaction potentials. The state decays through a
combination of dipolar spin-spin and second-order spin-orbit interactions
\cite{Mies:1996}. The wave function is expanded in a fully uncoupled basis set
that contains all allowed spin functions for each value of the end-over-end
angular momentum $L$ of the colliding pair.

We first perform calculations using a restricted basis set that includes only
functions with $L=0$. This removes channels with $M_F = m_{f\textrm{,a}} +
m_{f\textrm{,b}}>-4$, including the open channels, so that the quasibound
states of interest become truly bound. We locate these bound states using the
\textsc{bound} package, and use the results as initial estimates of the
resonance positions. We then switch to a full basis set, including functions
with $L=0$ and 2, and carry out scattering calculations using the
\textsc{molscat} package.

\begin{table}[tbp]
\caption{Convergence on the near-threshold state for two $^{85}$Rb atoms in
their $f=2, m_f=-2$ state at 155~G. The estimated value of
$\delta_\textrm{bg}=0.137\pi$ remains stable throughout the convergence. Units of
energy are $h\times$~Hz.
\label{table:Rb2}} \centering
\begin{ruledtabular}
\begin{tabular}{{c} *{2}{c} *{2}{c} }
 & & & \multicolumn{2}{ c }{ Estimated values} \\ \cline{4-5}
 $n$	& $E_n$			& $\delta/\pi$		& $E_\textrm{res}$	& $\Gamma$	\\ \hline
 $1$ & $-2056.301$ 		& $0.1366704$		& -		 		& -			\\
 $2$ & $-2035.942$ 		& $0.1366715$ 	& - 				& - 			\\
 $3$ & $-2015.582$ 		& $0.1366726$		& $-210.231$		& $1.138$		\\
 $4$ & $-210.231$ 		& $0.1389241$		& $-126.108$		& $1.246$		\\
 $5$ & $-126.108$		& $0.1261139$		& $-141.453$		& $1.010$		\\
 $6$ & $-141.453$		& $0.3103042$		& $-140.654$		& $0.969$		\\
 $7$ & $-140.654$		& $0.6353210$		& $-140.652$		& $0.971$		\\
 $8$ & $-140.749$		& $0.5739046$		& $-140.652$ 		& $0.971$		\\
 $9$ & $-139.681$		& $0.9891443$		& $-140.652$ 		& $0.971$		\\
\end{tabular}
\end{ruledtabular}
\end{table}

\begin{figure}[tbp]
\includegraphics[width=0.95\columnwidth]{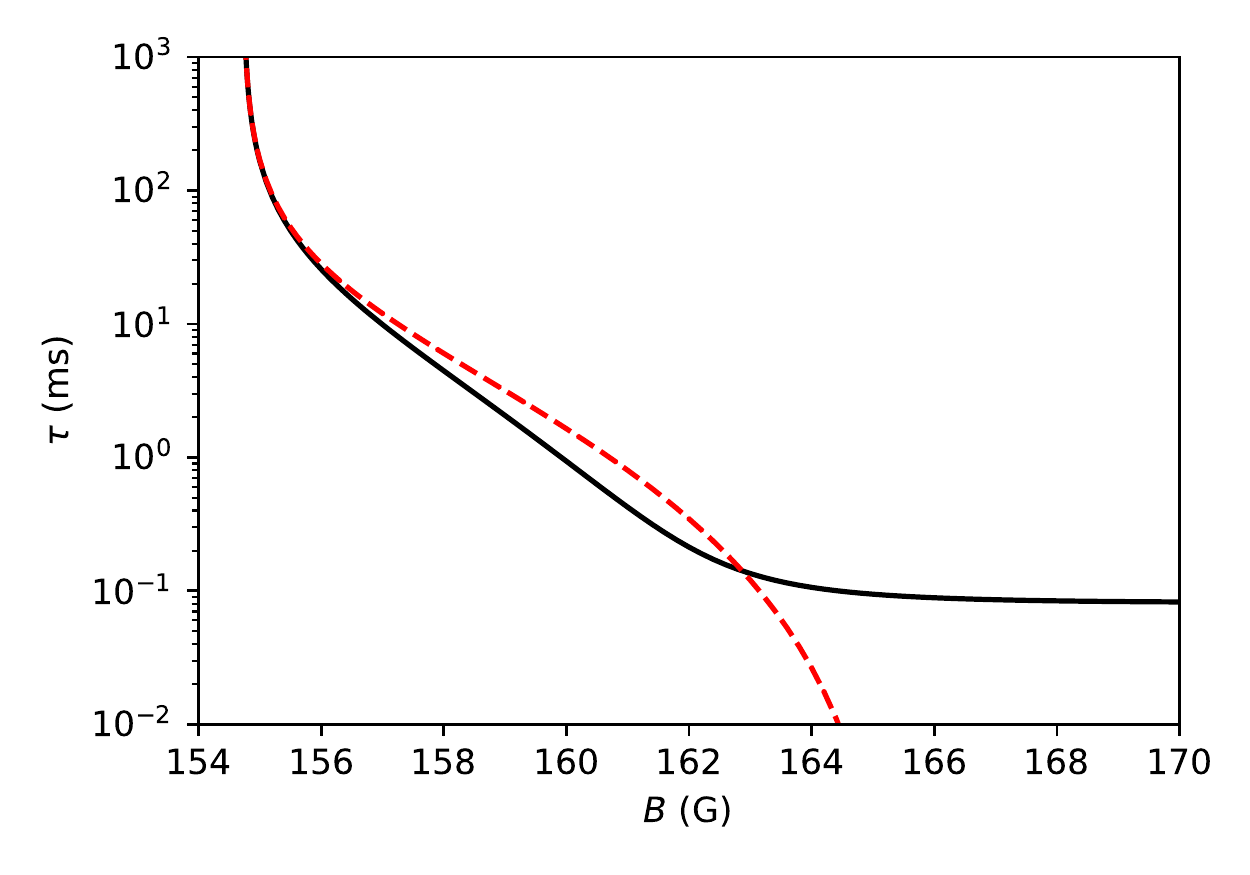}
\caption{Lifetime $\tau$ of the state below the $(f,m_f)$ = $(2,-2) + (2,-2)$ threshold as a
function of magnetic field near the 155~G resonance. The black line shows the lifetime obtained
using the present method and the red dashed line shows the result of Eq.\ (9) of
K\"ohler \emph{et al.}\ \cite{Kohler:2005}.} \label{fig:B_tau}
\end{figure}

Table \ref{table:Rb2} shows the convergence on the bound state below the
$(2,-2) + (2,-2)$ threshold at a magnetic field of $B=155$~G, where it is bound
by only $140\times h$~Hz and is less than $1\times h$~Hz wide. Under these
conditions, the initial estimate of the binding energy from bound-state
calculations is inaccurate by over 1900 times the width, which is about 13.5
times the binding energy. Nonetheless, our method converges with calculations
at only 9 energies. Remarkably, even the estimate of the width from the first 3
points is within 20\% of the converged value.

We repeat this convergence procedure for a range of magnetic fields and obtain
the lifetimes shown in Fig.\ \ref{fig:B_tau}. At fields below 155~G, where the
quasibound state is extremely narrow, the initial estimate of the energy from
bound-state calculations is inadequate. In this region we obtain the initial
estimate from linear extrapolation based on previous energies.

K\"ohler \emph{et al.}\ \cite{Kohler:2005} developed a model for the decay of this bound state.
They obtained an analytic expression for the lifetime at magnetic fields just above the resonance,
where the bound state is dominated by the open-channel component. Figure \ref{fig:B_tau} includes
the result of Eq.\ (9) of Ref.\ \cite{Kohler:2005}; we use values for the real and
imaginary parts of the scattering length from our coupled-channel calculations. The analytic
expression of Ref.\ \cite{Kohler:2005} is accurate close to the resonance, but breaks down far from
it. The coupled-channel results correctly approach the lifetime of the underlying quasibound state,
but the analytic expression for the lifetime drops unphysically to zero, because it is based on an
expression for the closed-channel amplitude that is valid only close to resonance.

K\"ohler \emph{et al.}\ performed coupled-channel scattering calculations at 9 specific values of
the magnetic field. The calculations were too laborious for them to show the results as a curve
rather than isolated points. It should be noted that they used an earlier interaction potential
than the present work, which produced a significantly shorter lifetime (32 $\mu$s) for the
underlying state. We have investigated the dependence of this lifetime on the dipolar spin-spin and
second-order spin-orbit interactions, which have opposite sign for the interaction potential of
Ref.\ \cite{Strauss:2010}. We find that the lifetime is strongly dependent on the degree of
cancelation between the two.

\subsection{Energy dependence of background scattering}

At each step, the procedure described here obtains estimates of the resonance position and width.
In their simplest form, these estimates neglect the energy dependence of the background phase shift
or eigenphase sum $\delta_\textrm{bg}(E)$. They are in error if $\delta_\textrm{bg}(E)$ varies
significantly across the current range of points. A background energy dependence may prevent
identification of the resonance from far away and, in drastic cases, may prevent convergence on the
resonance even from points near to it. However, in many cases it is possible to obtain an estimate
of $\delta_\textrm{bg}(E)$. It is then straightforward to redefine $\widetilde{a}(E)=\tan (\delta
(E)-\delta_\textrm{bg}(E))$ and use the rest of the procedure exactly as before.

The estimate of $\delta_\textrm{bg}(E)$ may come from a variety of sources. It may be as simple as
estimating the background gradient from a plot of $\delta (E)$, or it might include the effects of
other resonances whose properties are already known. More sophisticated estimates may also be
envisaged, based for example on semiclassical phase integrals for non-resonant open channels. In
any case it is not necessary for the estimate to include the \emph{constant} part of
$\delta_\textrm{bg}(E)$, which is accounted for by the 3-point approach, but only its energy
dependence.

In the presence of uncorrected background variation in
$\delta_\textrm{bg}(E)$, the resonance parameters obtained depend on the
positions of the outer two points within the range permitted by the
tolerance $\xi$. Different convergence sequences may result in
differently positioned points and produce small apparently random
variations in the parameters. Such variations may be unacceptable in
some applications, such as calculating numerical derivatives of
resonance parameters when performing least-squares fits of potential
parameters to experimental data. They may be reduced or eliminated by
choosing a small value of the tolerance $\xi$.

\section{Partial widths}

The product state distribution from decay of a quasibound state is
characterized by a set of partial widths $\Gamma_i$ for each open channel $i$.
For an isolated narrow resonance, the partial widths sum to the total width
$\Gamma$. Across a resonance, each S-matrix element describes a circle in the
complex plane \cite{Brenig:1959, Taylor:1972}
\begin{equation}
S_{ii'}(E)=S_{\textrm{bg},ii'}-\frac{\textrm{i}g_ig_{i'}}{E-E_\textrm{res}+\textrm{i}\Gamma/2}.
\label{eq:S_circle}
\end{equation}
The partial widths are defined as real quantities, $\Gamma_i=|g_i|^2$, and the
circles in the complex plane have radii $\sqrt{\Gamma_i\Gamma_{i'}}/\Gamma$;
the complex coefficients $g_i$ also have phases $\phi_i$, which determine the
orientations of the circles in the complex plane.

For each S-matrix element, Eq.\ \eqref{eq:S_circle} contains 6 independent
parameters. We can determine the magnitude and phase of the product $g_i
g_{i'}$, but not $g_i$ and $g_{i'}$ individually. As each element has real and
imaginary parts, we need calculations at 3 points to determine these 6
parameters. However, the position $E_\textrm{res}$ and total width $\Gamma$ are
common to the behavior of all S-matrix elements. Our usual procedure is
therefore to keep these two parameters fixed at the values determined from
fitting to the eigenphase sum, as described in Section \ref{sec:theory}, and
use calculations at only 2 points in order to characterize the real and
imaginary parts of $S_{\textrm{bg},ii'}$ and $g_i g_{i'}$.

Equation \eqref{eq:S_circle} may be written
\begin{equation}
S(E)=S_\textrm{bg}+\frac{\textrm{i}D}{x+\textrm{i}}, \label{eq:S_circle_red}
\end{equation}
where $x=2(E-E_\textrm{res})/\Gamma$ and $D$ is a matrix with elements
$D_{ii'}=-2g_i g_{i'}/\Gamma$. We insert calculated S matrices $S_1$ and $S_2$
at two energies $E_1$ and $E_2$, and solve the resulting simultaneous equations
to give
\begin{align}
S_\textrm{bg}&=\frac{(x_1+\textrm{i})S_1-(x_2+\textrm{i})S_2}{x_1-x_2};\\
D&=(1-\textrm{i}x_1)(S_1-S_\textrm{bg}).
\end{align}
The partial widths are obtained from the diagonal elements,
$\Gamma_i=|D_{ii}|\Gamma/2$.

We evaluate the partial widths after convergence on $E_\textrm{res}$ and
$\Gamma$ as described in Section \ref{sec:theory}. We use S matrices at two of
the three energies used in the final iteration of the convergence procedure;
these S matrices have already been calculated during the convergence, so we
need no additional coupled-channel calculations. For narrow resonances
(including the examples covered in this paper), where convergence is limited by
numerical noise, it is best to use the points closest to and furthest from the
resonance position. For wide resonances, where convergence is limited by a
non-constant background, it is best to use the two points closest to the
resonance position.

\begin{table}[tbp]
\caption{Partial widths for the $v=1$, $j=0$ state of Ar--H$_2$. All decay channels have $v=0$ and
are labeled by their rotational quantum number $j$. Units of energy are $10^{-10}\times
hc$~cm$^{-1}$. \label{table:ArH2_partial}} \centering
\begin{ruledtabular}
\begin{tabular}{cr@{\!\!\!\!}lr@{\!\!\!\!}lcc}
 $j$	& \multicolumn{2}{ c }{$S_{\textrm{bg},jj}$}	& \multicolumn{2}{ c }{$D_{jj}$}			& \multicolumn{2}{ c }{$\Gamma_j$}						 \\ \cline{6-7}
 	&  			&						&			&					& this work			& Ref.\ \cite{Hutson:ArH2:1983} 	\\ \hline
 0	& $-0.741$	& ${}+ 0.083 \textrm{i}$		& $-0.017$	& ${}+ 0.009 \textrm{i}$	& $2.0$				& $2.0$	\\
 2	& $0.623$		& ${}- 0.229 \textrm{i}$		& $0.003$		& ${}+ 0.183 \textrm{i}$	& $18.5$				& $18.5$	\\
 4	& $-0.831$	& ${}+ 0.367 \textrm{i}$		& $1.203$		& ${}- 0.614 \textrm{i}$	& $136.9$				& $136$	\\
 6	& $-0.454$	& ${}+ 0.881 \textrm{i}$		& $0.294$		& ${}- 0.337 \textrm{i}$	& $45.3$				& $45$	\\
 8	& $-0.316$	& ${}+ 0.949 \textrm{i}$		& \multicolumn{2}{ c }{$\sim 10^{-6}$}	& $0.0002$			& $<0.01$
\end{tabular}
\end{ruledtabular}
\end{table}

We have applied this procedure to the Ar--H$_2$ example described above, and the
results are shown in Table \ref{table:ArH2_partial}. It may be seen that they are
in excellent agreement with those of ref.\ \cite{Hutson:ArH2:1983}, which were
obtained by a much more laborious fitting procedure.

It is also possible to use a 3-point approach based on the resonant circle in a
single S-matrix element. This is analogous to the ``fully complex" procedure of
Ref.\ \cite{Frye:resonance:2017} and allows convergence on $E_\textrm{res}$ and
$\Gamma$ as well as $S_{\textrm{bg},ii'}$ and $D_{ii'}$. In multichannel
scattering, we have sometimes found that this procedure converges successfully,
even for resonances where uncorrected energy dependence in the background
prevents convergence based on the eigenphase sum. This occurs because the
eigenphase sum contains background contributions from \emph{all} open channels,
whereas a diagonal S-matrix element contains a background contribution from
only a single open channel. Nevertheless, it is usually more
convenient to locate resonances initially in the eigenphase sum, using an
estimated background energy dependence if necessary.

\smallskip
\section{Conclusions}

We have developed an automated procedure to converge on a scattering resonance
and characterize it in terms of the resonance energy and width and the phase
shift (or S-matrix eigenphase sum) of the background scattering. The procedure
may be used for both single-channel and multichannel scattering. It eliminates
the extensive manual iteration and fitting inherent in previous approaches.

The procedure requires an initial set of three energies in the vicinity of the
resonance, one of which must be close enough to the resonance that its phase
shift is affected by the resonance by more than the variation of the background
phase across the range of the points. The strategy employed aims to generate a
set of three points, one very close to the center of the resonance and two
others at varying distances on either side. The three points allow
characterization of the resonance position, width and background phase.
Subsequent processing of the S-matrix elements allows extraction of partial
widths for decay to individual open channels.

We have demonstrated the method on two very different test cases. One is a very narrow state in the
Van der Waals complex Ar--H$_2$, which lies far from any threshold and decays only by vibrational
predissociation. The other is a near-threshold state of $^{85}$Rb$_2$, whose lifetime varies from
$\sim$80~$\mu$s to over 1~s as a function of magnetic field.

An important use of the procedure will be in least-squares fitting to determine
interaction potentials from experimental results \cite{H92ArHF, I-NoLLS,
Takekoshi:RbCs:2012, Berninger:Cs2:2013, Julienne:Li67:2014}. Such fits require
numerical derivatives of calculated properties with respect to potential
parameters, and the calculation of these derivatives is not feasible if manual
intervention is required.

We have implemented the procedure in version 2020.0 of the general-purpose quantum scattering package \textsc{molscat} \cite{mbf-github:2020}.

\begin{acknowledgments}
We are grateful to Paul Julienne and Ruth Le Sueur for valuable discussions.
This work was supported by the U.K. Engineering and Physical Sciences Research
Council (EPSRC) Grants No.\ EP/N007085/1, EP/P008275/1 and EP/P01058X/1.
\end{acknowledgments}

\bibliography{../all}

\end{document}